\documentclass[11pt,a4paper]{article}

\usepackage{epsfig}
\usepackage{latexsym}
\usepackage{amsfonts}
\usepackage{amsmath}
\usepackage{amsthm}
\usepackage{amssymb}
\usepackage{amsbsy}
\usepackage{multirow}
\usepackage{slashed}
\usepackage{color}
\usepackage{mathrsfs}
\usepackage{subfigure}
\usepackage[font={footnotesize,sl},bf]{caption}
\usepackage{wrapfig}
\usepackage{jheppub}

\newcommand\be{\begin{equation}}
\newcommand\bea{\begin{eqnarray}}
\newcommand\ee{\end{equation}}
\newcommand\eea{\end{eqnarray}}

\newcommand{\bdm}{\begin{displaymath}}
\newcommand{\edm}{\end{displaymath}}
\newcommand{\nn}{\nonumber \\}
\newcommand{\f}[2]{\frac{#1}{#2}}
\newcommand{\bref}[1]{(\ref{#1})}

\title{Black rings with fourth dipole cause less hair loss}

\author{Borun D.\ Chowdhury}
\affiliation
{Institute for Theoretical Physics, University of Amsterdam,\\
Science Park 904, Postbus 94485, 1090 GL Amsterdam, The Netherlands}

\abstract{An example of entropy enigma with a controlled CFT dual was recently studied in \cite{Bena:2011zw}. The enigmatic bulk configurations, considered within the STU model, can be mapped under spectral flow into black rings with three monopole and dipole charges. Even though the bulk and CFT configurations existed in the same region of parameter space, the Bekenstein-Hawking entropy of the bulk configurations was found to be lower than the microscopic entropy from the CFT. While it is possible that the difference in entropy is due to the bulk and boundary configurations being at different points in the moduli space, it is also possible that the bulk configurations embeddable within the STU model are not the most entropic. New families of BPS black ring solutions with four electric and four dipole magnetic charges have recently been explicitly constructed in \cite{Giusto:2012gt}. These black rings are not embeddable within the STU model. In this paper we investigate if these black rings can be entropically dominant over the STU model black rings. We find that the new black rings are always entropically subdominant to the STU-model black rings. However, for small fourth dipole charge these black rings continue to be dominant over the BMPV in a small region of parameters and are thus enigmatic.}

\begin{document}
\maketitle

\section{Introduction}

Recently there has been a great interest in studying black holes with a condensate of some field outside the horizon \cite{Gubser:2008px,Hartnoll:2008vx,Hartnoll:2008kx,Denef:2009tp,Gubser:2009qm,Gauntlett:2009dn,Gauntlett:2009bh,Correa:2010hf,Basu:2010uz,Bhattacharyya:2010yg,Dias:2011tj,Alsup:2011qn,Anabalon:2012sn,Brihaye:2012cb,Capela:2012uk}\footnote{Many authors refer to this condensate as hair of black hole. However,  originally this term was used to describe microstates of black holes and not a condensate outside the horizon. Since the term has gained acceptance we will use it to refer to condensates outside horizons.}. Often there is also a black hole with the same charges as the black hole with a condensate outside but with a lower Bekenstein Hawking entropy. Such cases can be thought of as the black hole emitting some charges to increase its entropy. These thermodynamic instabilities are often associated with super-radiance \cite{Basu:2010uz,Chowdhury:2010ct}
which is quite like the process of a population inverted gain medium emitting photons in a LASER. 
Entropy enigmas \cite{Gauntlett:2004wh,Denef:2007vg,deBoer:2008fk,Bena:2011zw} are also examples of this phenomenon where certain two center configurations can have a larger entropy than a single center configuration. In \cite{Anninos:2011vn,Chowdhury:2011qu} non-extremal two center configurations were studied with one of the centers as a probe. These systems exhibited a thermodynamic instability of the black hole towards spitting out massive non-perturbative objects. These can be thought of as a probe analysis of the entropy enigma for non extremal black holes.

\begin{figure}[tb]
 \begin{center}
\begin{tabular}{cc}
 \epsfxsize=8cm \epsfbox{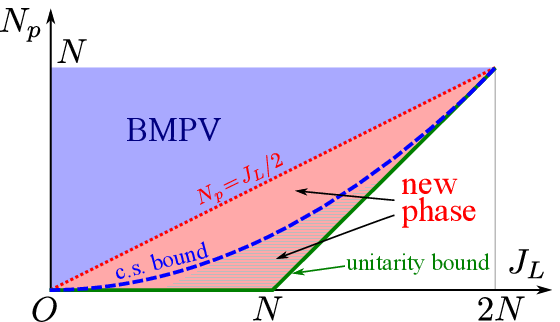} 
 &
 \epsfxsize=8cm \epsfbox{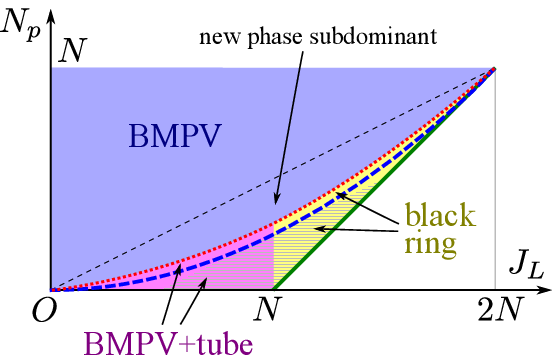}  \\[-25.8cm] 
 \begin{minipage}{8cm}(a) CFT phase diagram in the RR sector\\
 at the orbifold point\end{minipage}
 &
 (b) Bulk phase diagram 
\end{tabular}
\caption{\sl The updated, correct phase diagram of the D1-D5 system for
the CFT and bulk (schematic, not to scale). The abbreviation ``c.s.\
bound'' refers to the cosmic censorship bound $N_p=J_L^2/4N$.  
  %
  \label{fig:new_phase_diag}}
\end{center}
\vspace*{-.5cm}
\end{figure}

In \cite{Bena:2011zw} a family of novel supersymmetric phases of the D1-D5 CFT at the orbifold point was found which, in certain ranges of parameters, have more entropy than all other known ensembles. Dual bulk configurations in the same ranges of parameters were also found which, for  smaller ranges of parameters, have more entropy than a BMPV black hole\cite{Breckenridge:1996is}. These solutions were found at the point in the moduli space where none of the moduli are turned on and supergravity is weakly coupled\cite{Seiberg:1999xz}. For the rest of the paper we will refer to this simply as the supergravity point unless specified otherwise. The entropy of the bulk configuration was found to be smaller than that of the CFT phase indicating that some of the CFT states lift when moving away from the orbifold point. These configurations were the first instance of the black hole entropy enigma\cite{Gauntlett:2004wh,Denef:2007vg,deBoer:2008fk} with a controlled CFT dual. Furthermore, contrary to common lore these objects existed outside the cosmic censorship bound\footnote{In \cite{deBoer:1998us,Maldacena:1999bp} it was shown that the (modified) elliptic genus calculated from the CFT and from counting supergravity particles, excluding black holes, matches in most of the region outside the cosmic censorship bound (in the hatched regions in figure \ref{fig:new_phase_diag}). It is usually believed that at strong coupling anything not protected, i.e. not contributing to the (modified) elliptic genus, will lift and not contribute to the partition function. Thus this resulted in the common lore that there are no black objects beyond the cosmic censorship bound. The {\em cosmic censorship bound}, in light of results of \cite{Bena:2011zw}, is now understood to be the bound on the single center black hole.}.

The phase diagram at the orbifold point and the supergravity point are reproduced from \cite{Bena:2011zw} in figure \ref{fig:new_phase_diag}. Here $N_p$ is the amount of  momentum charge, $N$ is the product of the number of D1 and D5 branes and $J_L, J_R$ are the angular momenta. Figure \ref{fig:new_phase_diag}a is the phase diagram at the orbifold point. In the Cardy regime $N_p - J_L^2/4 N \gg N$ the long string sector, which is dual to the BMPV black hole, dominates. This phase has $J_R=0$. However, outside the Cardy regime the new phase of the CFT comes into existence, and dominates, in the region $J_L>2 N_p$ all the way to the unitarity bound. This phase has $J_R = J_L -2 N_p$. Figure \ref{fig:new_phase_diag}b is the phase diagram at the supergravity point. In the Cardy regime the BMPV black hole dominates. Since we are looking for duals to the CFT phase we focus on configurations which have $J_R = J_L -2 N_p$. There are two such dominant configurations viz. a bound state of BMPV and supertube, and a black ring. These are related by spectral flow\cite{Bena:2008wt}. For $J_L<N$ the former dominates entropically while in the rest of the region the latter dominates. Both of these exist all the way to the unitarity bound. However, they dominate over the BMPV in a smaller region than the corresponding phase at the orbifold point. Neither the CFT new phase nor the bulk configurations contribute to the elliptic genus\cite{Dabholkar:2009dq,Bena:2011zw}. 

Supersymmetric states which do not contribute to the elliptic genus can become non-supersymmetric when moving in the moduli space. If this happens then the dimension of such states is not protected and then they do not contribute to the partition function. This could explain the mismatch between the entropies at the two points in the moduli space considered above. It could be possible that some states become non-supersymmetric and heavy on moving away from the orbifold point while the others do not. This is an interesting possibility which suggests that the entropy might further decrease {\em or maybe even increase} on turning on various moduli. The moduli required to move away from the supergravity point considered in \cite{Bena:2011zw}  involve the self dual NS-NS B field on $T^4$ and a linear combination of $C^{(0)}$ and $C^{(4)}$ Ramond-Ramond field in the D1-D5 frame\cite{Dhar:1999ax}. These moduli cannot be turned on in the STU model\footnote{We thank Masaki Shigemori for pointing this out to us.}.

There is another, more mundane, possibility which can explain the mismatch between the entropies at the two points in the moduli space. It is  possible that the STU truncation does not allow the maximum entropy configuration. It would then be interesting to turn on more fields than allowed by the STU truncation while keeping the same moduli. 
Motivation to consider more fields has been provided by a recent CFT analysis of D1-D5-P bound states\cite{Giusto:2011fy}.  Following this motivation,
in \cite{Giusto:2012gt}  a new black ring, not contained in the STU truncation, was constructed. It has an extra charge and dipole in comparison to the black ring in the STU model \cite{Elvang:2004rt}. It is a solution of $\mathcal N=2$ supergavity with 3 vector multiplets (the STU model having only two). Three is the maximum number of vector multiplets one can turn on while preserving the rotation symmetry of $T^4$ in type IIB duality frame. It is possible to turn off the fourth charge in these black rings while keeping the fourth dipole charge. It is then natural to ask if these black rings could be the dominant saddle point dual to the enigmatic phase found in \cite{Bena:2011zw}. In this paper we answer this question and find that the entropy of this new black ring is always subdominant to the STU black ring. However, these black rings, for small values of the fourth dipole charge, continue to exist outside the cosmic censorship bound. In addition for small values of the fourth dipole charge these black rings are dominant over the BMPV black hole in some small region of parameter space. Thus these black rings are enigmatic also.

There is a lore that addition of charges leads to decrease in entropy of black objects and one might wonder why our result was not obvious from the beginning. While it is true that in most cases this lore is true, there are counter examples. In fact \cite{Bena:2011zw} showed that in certain regimes of parameters a black hole is entropically subdominant to a black ring which has more charges turned on. Thus our result is not at all obvious from the beginning. Indeed the fact that in certain regimes the black ring with fourth dipole is still entropically dominant over the black hole shows that even for the fourth dipole it is not always true that addition of charge decreases entropy.

The organization of this paper is as follows. In section \ref{Review} we review the enigmatic phase at the orbifold point in the CFT and the  dual saddle points  at the supergravity point within the STU model. In section \ref{NewBR} we review the essential properties of new black rings and find the maximum entropy configurations. In section \ref{Conclusion} we conclude with a discussion and possible future directions.

\section{Review of moulting black holes} \label{Review}

In this section we review the CFT construction of the new phase at the oribifold point and the bulk construction of possible duals at the supergravity point from \cite{Bena:2011zw}. We assume some familiarity with the D1-D5 system at the orbifold point. For details of the D1-D5 CFT the reader is referred  to \cite{Avery:2009tu} for example.

\subsection{The new phase at the orbifold point in the D1-D5 CFT} \label{ReviewCFT}

The Higgs branch of the D1-D5 system flows in the IR to $1+1$ dimensional $\mathcal N =(4,4)$ superconformal field theory whose target space is a resolution of the orbifold $Sym^{N}(M)$, where we have defined $N \equiv N_1 N_5$ with $N_1, N_5$ being the number of D1 and D5 branes respectively. While $M$ can be either $T^4$ or $K3$ \cite{David:2002wn}, we focus on $T^4$ for specificity.  The existence of the new phase is independent of this choice. The orbifold point is the point in moduli space where the target space is $Sym^{N}(M)$.

\begin{figure}
 \begin{center}
 \begin{tabular}{ccc}
   \epsfxsize=5cm \epsfbox{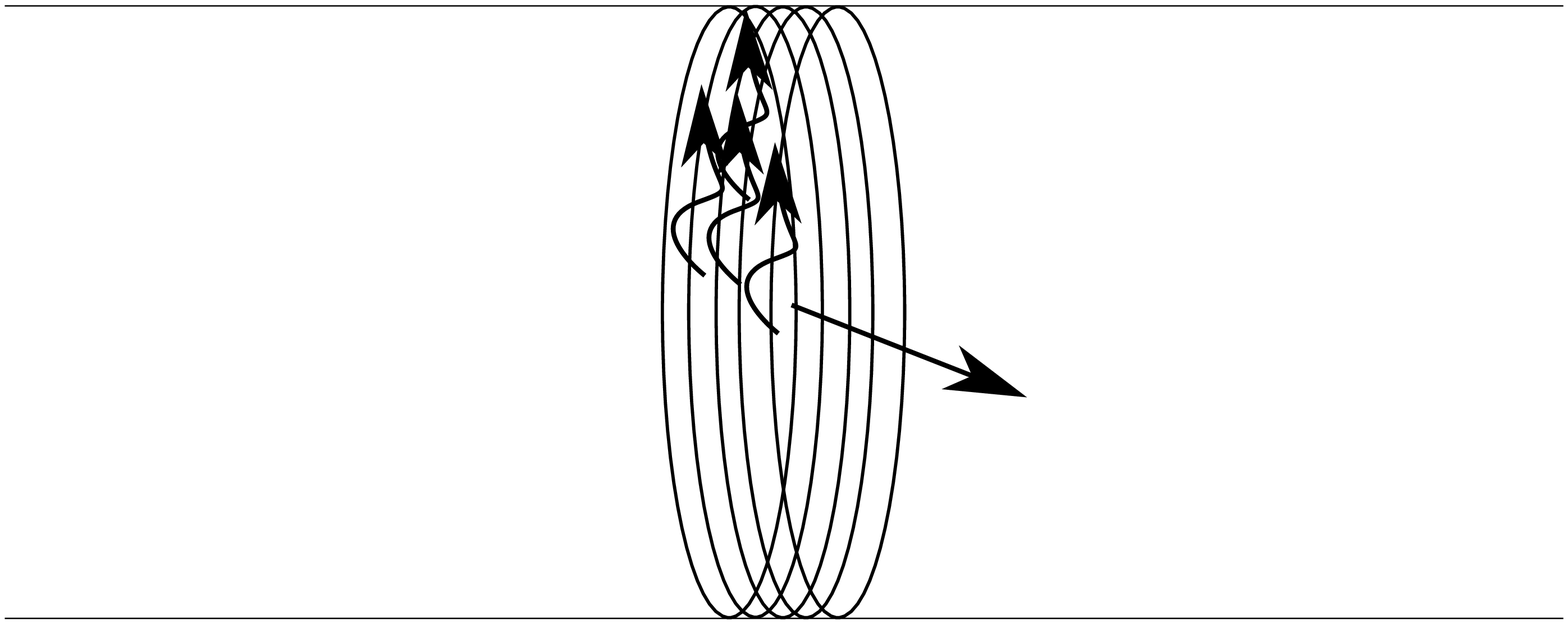} &
   \epsfxsize=5cm \epsfbox{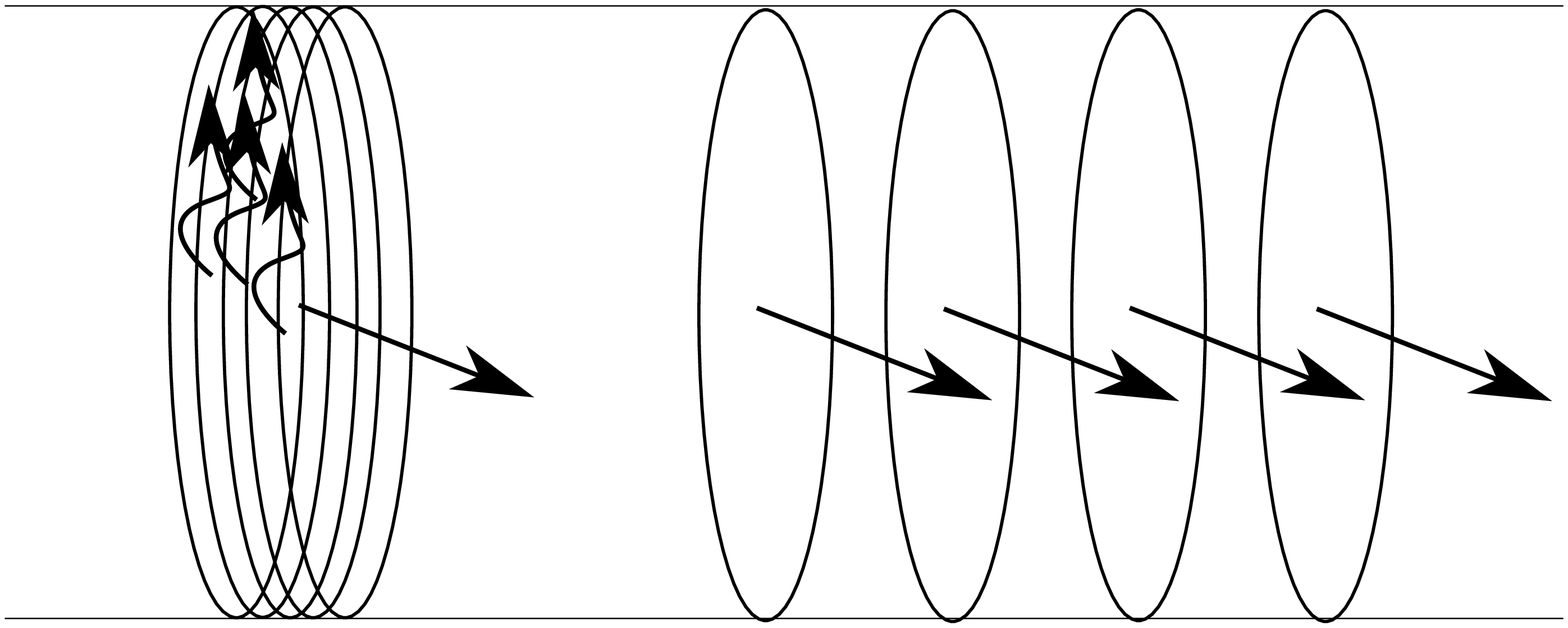}\\
 (a)  BMPV &
 (b) Enigmatic Phase
 \end{tabular}\\[2ex]
 \caption{\sl The long string phase (dual to the  BMPV black hole) and the enigmatic phase at the orbifold point of the D1-D5 CFT.\label{fig:3phases}}
 \end{center}
\end{figure}

The orbifolding produces various winding sectors with the total winding fixed to be $N$. The central charge is given by $c=6N$. The Ramond-Ramond sector ground states have weight $(\f{N}{4}, \f{N}{4})$ in the left and right sector. The momentum is identified with  excitations in the CFT. The bosonic excitations are uncharged while the fermionic excitations are charged under the $SU(2)_L \times SU(2)_R$ R-symmetry.
States with only left moving  (or equivalently only right moving) momentum turned on are supersymmetric and the total dimension is given by $L_0 = N_p + \f{N}{4}$. In the Cardy regime $N_p - \f{J_L^2}{4N} \gg N$ and the entropy is given by
\be
S_{cardy} = 2 \pi \sqrt{N N_p - \f{J_L^2}{4}}
\ee
which is reproduced to leading order by having the CFT in the completely twisted sector with all the excitations living on  this long string as shown in figure \ref{fig:3phases}a. All the macroscopic angular momentum is carried by fermionic excitations on the long string.  Thus, the BMPV black hole is understood to be dual to the completely twisted sector in the CFT. Both the BMPV black hole and the long string sector exist only above the cosmic censorship bound.

Outside the Cardy regime the entropy depends on the details of the CFT and one could look for a sector at the orbifold point where $l$ winding, out of the total $N$, is carried by short strings whose angular momenta, coming from fermion zero modes, are aligned. This sector is depicted in figure \ref{fig:3phases}b. As shown in \cite{Bena:2011zw}, the entropy of this sector is maximized for $l=J_L-2 N_p$ and is given by 
\be
S_{new} = 2 \pi \sqrt{N_p (N+N_p- J_L)}. \label{CFTEntropy}
\ee
This phase exists in the orange triangle shown in figure \ref{fig:new_phase_diag}a and its entropy dominates over the BMPV in the coexistence region. This phase, unlike the long string sector, has a non-zero $J_R$ given by
\be
J_R= J_L-2 N_p \label{Eqn:JRconstraint}
\ee
coming from the $SU(2)_R$ R-charge carried by the fermion zero modes on the short strings. Thus this phase transition can be viewed as a thermodynamic instability in a mixed ensemble with $L_0$ and $J_L$ held fixed but $J_R$ not fixed.
\begin{figure}[htb]
\begin{center}
   \epsfxsize=12cm \epsfbox{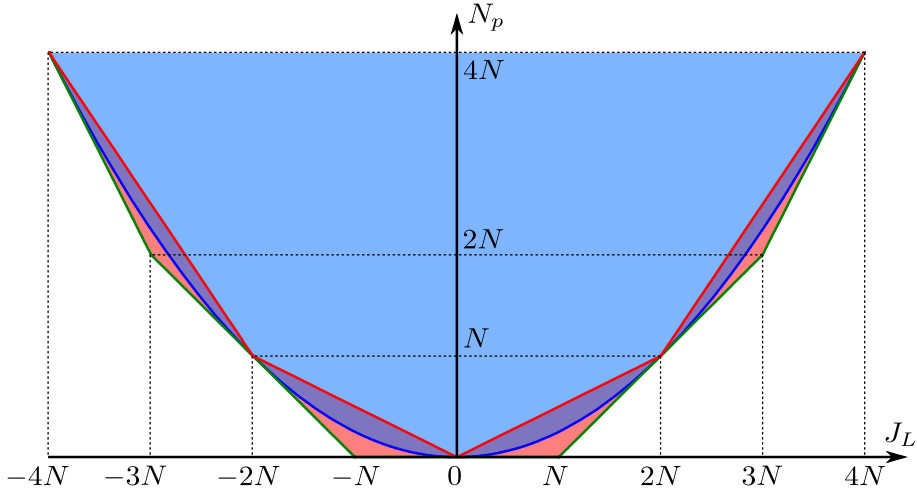}
\end{center}
\caption{\sl Spectral-flowed enigmatic phases.}  \label{fig:spectralFlowedEnigmaticPhase}
\end{figure}
This enigmatic phase can be mapped to other enigmatic phases under spectral flow
\bea
N_p \to N_p - k  J_L + k^2 N, \qquad J_L \to J_L - 2 k N \label{Eqn:SF}.
\eea	
Here $k \in \mathbb Z$ ensures the fermion periodicity around the field theory direction is maintained. Spectral flow maps the triangular region of existence to other triangles. These regions are shown in figure \ref{fig:spectralFlowedEnigmaticPhase}.  We refer to the region  bounded by
\begin{gather}
N_p - k J_L +k^2 N>0, \nn
N_p - (k+1) J_L +(k+1)^2 N > 0, \nn
J_L(1+ 2 k)  - 2 N_p-2 k (1+k)N >0 \label{kthwedge}
\end{gather}
as the $k^{th}$ wedge and the corresponding phase as the $k^{th}$ phase. The  entropy of the $k^{th}$ phase is given by
\be
S_{k}=2 \pi \sqrt{(N_p - k J_L +k^2 N)(N_p - (k+1) J_L +(k+1)^2 N)}.
\ee

\subsection{The new phase at the supergravity point in the STU model}

It was shown in \cite{Bena:2011zw} that all two center bulk configurations with one center smooth and the other black, contained within the STU model, can be related by spectral flow to a black ring\footnote{More correctly \cite{Bena:2011zw} showed that two center configurations with one center smooth and the other black can be spectral flowed to a supertube bound to a BMPV black hole. However, this configuration itself can be spectral flowed to a black ring.}. Thus, the maximization of entropy can be done by spectral flowing to the black ring, maximizing the entropy and flowing back. 

We will now look at black rings embeddable within the STU truncation which are asymptotically $AdS_3 \times S^3$. The procedure for their construction in terms of harmonic functions on a Gibbons Hawking base is well established and we refer the reader to \cite{Bena:2007kg, Bena:2011zw} for details. The decoupling limit is explained in \cite{Bena:2008wt,Bena:2011zw} for instance. The black ring is obtained by compactifying M-theory on $T^2 \times T^2 \times T^2$ and its charges and dipole charges come from M2, M5 branes wrapping these 2-tori. Concretely, the black ring is specified in terms of three monopole charges $\bar N_i$ coming from $M2$ branes wrapping the three $T^2$, three dipole charges $n_i$ coming from $M5$ branes wrapping the three $T^2 \times T^2$ and a closed loop in the non-compact direction labeled by $\psi$, and two angular momenta $J_L, J_R$. Compactifying on one direction from the first $T^2$ and T dualizing along the other, followed by a further T-duality on both directions of the second $T^2$, we obtain the black ring in type IIB. Denoting the remaining direction of the first $T^2$ as $S^1$  we have the compactification $S^1 \times T^2 \times T^2$. In this frame the charges are momentum and D1 along $S^1$ and D5 along $S^1 \times T^2 \times T^2$. The dipole charges are $D1$ wrapping $\psi$, D5 wrapping $\psi \times T^2 \times T^2$ and KK monopole along $\psi \times T^2 \times T^2$ with its fibre direction along $S^1$.

By the AdS/CFT duality, in the decoupling limit this configuration is dual to states in the D1-D5 CFT. However, since the D1-D5 CFT is the infrared fixed point of the theory of $D1$ branes wrapping $S^1$ and D5 branes wrapping $S^1 \times T^4$, the configurations described above fail to capture all the states of the D1-D5 CFT.

The total charges measured at infinity, $N_i$, include contributions from local monopole and dipole charges and are given by
\be
N_1 = \bar N_1 + n_2 n_3, ~~ N_2 = \bar N_2 + n_1 n_3, ~~ N_3 = \bar N_3 + n_1 n_2.
\ee
Quantization of flux enables us to normalize the charges to have $N_i, \bar N_i, n_i \in \mathbb Z$ \cite{Bena:2011zw} and the absence of closed timelike curves (CTCs) further constrains these numbers to 
\be
n_i, \bar N_i \ge 0. \label{CTCAbsenceCondition1}
\ee
The entropy of the black ring is given by
\be
S = 2 \pi \sqrt{D} 
\ee
where\footnote{Usually the formulae for black ring do not have the absolute value for the angular momenta written explicitly. However, since under spectral flow the angular momentum can change sign we write the absolute value explicitly for $J_L$. There is no such issue for $J_R$ so we keep the absolute value sign implicit.}
\be
D= n_1 n_2 (N_1 N_2 -n_3(|J_L|-n_3 N_3)) - \f{1}{4} (|J_L| -J_R - 2 n_3 N_3)^2. \label{STUBREntropy}
\ee
There is a further constraint on the black ring solution
\be
|J_L| - J_R  = \bar N_1 n_1 + \bar N_2 n_2 + \bar N_3 n_3 + 2 n_1 n_2 n_3 \label{Eq:STU-contraint}
\ee
relating the two angular momenta.
We are comparing the black ring entropy at the supergravity point with the new ensemble entropy at the orbifold point which only sees the combination $N= N_1 N_2$. Thus for simplicity  we specialize to the case
\be
N_1 = N_2 = \sqrt{N},  \qquad n_1=n_2 = \sqrt{n}.
\ee
The natural question to ask is which wedge in figure \ref{fig:spectralFlowedEnigmaticPhase} is the black ring dual to?
 We will first analyze this for  $J_L>0$, i.e. for $k \ge 0$, and then use the $J_L \to -J_L$ symmetry in \bref{STUBREntropy} to deduce the existence for $k<0$. If we take the black ring to be dual to the $k^{th}$ wedge  we  get the additional constraint 
\be
J_R =J_L(2k+1)  - 2 N_p-2 k (k+1)N \label{Eqn:JrKWedge}
\ee
from spectral flowing the $0^{th}$ wedge CFT constraint \bref{Eqn:JRconstraint} to the $k^{th}$ wedge. Inserting \bref{Eqn:JrKWedge} in \bref{Eq:STU-contraint} and solving for $\sqrt{n}$ we get
\be
\sqrt{n} = \frac{\sqrt{N}-\sqrt{2 n_3 k  ( J_L - (k+1) N)
 +N+n_3^2 N_3-2
   n_3 N_3}}{n_3}
\ee
where the other root is dropped by the requirement $\sqrt{N}>0$ to avoid CTCs \bref{CTCAbsenceCondition1}. Avoiding CTCs also requires $\sqrt{n}>0$ giving
\be
(2-n_3) N_3> 2 k (J_L - (k+1) N).   \label{STUNoCTC}
\ee
It is easy to see using \bref{kthwedge} that this condition can only be met if $n_3=1$ and  $k=0$ or $k=1$.  Thus the black ring with constraint \bref{Eqn:JrKWedge} can only exist in wedges $0$ and $1$ for $k \ge 0$.  By  the $J_L \to - J_L$  symmetry in \bref{STUBREntropy}  it can also exist only in  wedges $-1$ and $-2$ for $k<0$. The entropies in these wedges are given by
\begin{gather}
D_0 =  (N_p - J_L + N)( \sqrt{N} - \sqrt{N- N_p})^2, \\
D_1 = (N_p - J_L + N)( \sqrt{N} - \sqrt{2 (J_L- 2 N) + N- N_p})^2 - (J_L-2 N)^2, \\
D_{-1} =  \sqrt{N_p + J_L + N} (\sqrt{N} - \sqrt{N-N_p}), \\
D_{-2} = (N_p + J_L + N)( \sqrt{N} - \sqrt{2 (-J_L- 2 N) + N- N_p})^2 - (-J_L-2 N)^2.
\end{gather}
Here we have written $N_3$ as $N_p$ to emphasize that the third charge is momentum in the type IIB frame and to compare with the CFT variables.

\begin{figure}[htb]
 \begin{center}
   \epsfxsize=12.5cm \epsfbox{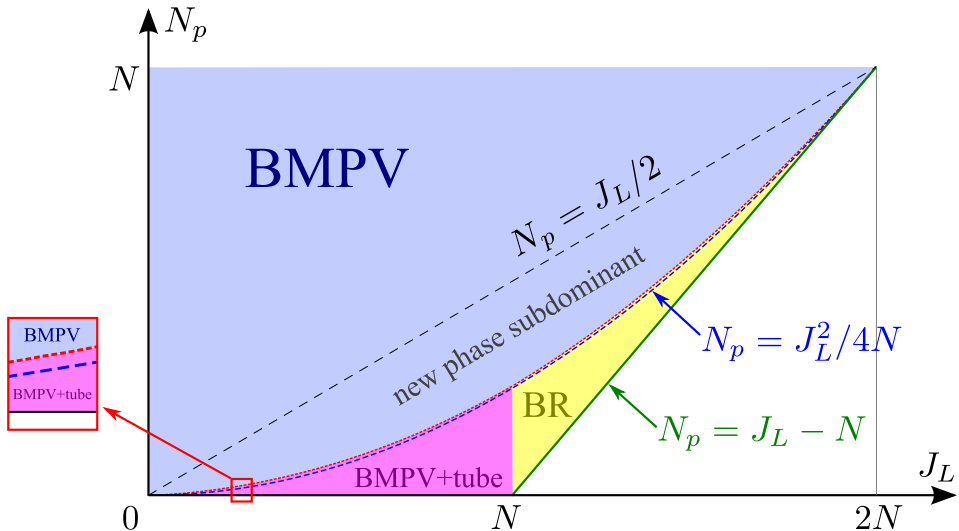} \caption{\sl The bulk
  phase diagram. In the light blue region, the single-center BMPV black
  hole is dominant.  In the pink and yellow regions the
  new phase dominates, either as a BMPV black
  hole surrounded by a supertube for $J_L<N$ (pink), or as a black ring
  for $J_L>N$ (yellow).  Below the thin dashed black line and above the
  dotted red curve, the BMPV phase and the new phase coexist but the BMPV phase is
  dominant.  In the narrow region between the dotted red curve and
  dashed blue curve, the two phases coexist and the new phase is
  dominant.
  \label{fig:phase_diag_grav}}
\end{center}
\end{figure}

Since we want to compare the configurations to the CFT states in the $0^{th}$ wedge we flow these expressions to the $0^{th}$ wedge using \bref{Eqn:SF}
\begin{align}
D_0 =  (N_p - J_L + N)( \sqrt{N} - \sqrt{N- N_p})^2 &\qquad  k=0, \\
D_{-1 \to 0} =  N_p (\sqrt{N} - \sqrt{J_L - N_p})^2  &\qquad k=1, \\
D_{1 \to 0} = D_{-1 \to 0}  - J_L^2  &\qquad k=-1, \\
D_{-2 \to 0} =  D_0 - (J_L-2N)^2 &\qquad k=2.
\end{align}
It is easy to see that in the first wedge for $J_L>N$ the maximum entropy is for $D_0$ while for $J_L<N$ it is $D_{-1 \to 0}$. In \cite{Bena:2011zw} it was shown that the latter is a bound state of a BMPV and supertube. The resulting phase diagram is shown in figure \ref{fig:phase_diag_grav}.  These configurations have lower entropy than that found at the orbifold point \bref{CFTEntropy}. However, in a small region of parameter space the entropy of these configurations is more than that of the BMPV black hole, so this is an example of entropy enigma. Since both the CFT configuration and the bulk configurations mentioned above are not counted by the elliptic genus, the mismatch in the entropy is no contradiction. However, it is interesting to observe that while some of the  states which are not protected do indeed lift, not all of them do.

\section{New black ring with entropy maximized} \label{NewBR}

In this section we will analyze $\mathcal N=2$ configurations with three vector multiplets studied in \cite{Giusto:2012gt} for possible duals to the dominant new CFT phase found in  \cite{Bena:2011zw}. These configurations have a extra charge and an extra dipole charge compared to the STU model. Since we want to study configurations with just D1, D5 and momentum charges, we will turn off the fourth charge, but keep the fourth dipole charge.

Just like in the previous section we will focus on black rings. We will see below that in the limit of large charges the black ring with the fourth dipole charge turned on is continuously connected to the black ring of the STU model.  Thus, it seems reasonable that all two center configurations with one center smooth and the other black are mapped by spectral flow to the black ring in the presence of the fourth dipole also, just like in the absence of the same. To prove this one would require generalization of the bulk spectral flows considered in \cite{Bena:2008wt} for the fourth dipole.

The black ring found  in  \cite{Giusto:2012gt} is specified by the three monopole and dipole charge of the STU model, the two angular momenta and in addition by one extra monopole and dipole charge $\bar N_4$ and $n_4$. In the type II B frame the charge corresponds to an equal amount of $NS1$ and $NS5$ branes wrapping $S^1$ and $S^1 \times T^2 \times T^2$ and the dipole to an equal amount of $NS1$ and $NS5$ branes wrapping  $\psi$ and $\psi \times T^2 \times T^2$, where $\psi$ is a loop in the non-compact directions.

As in the previous section we look at asymptotically $AdS_3 \times S^3$ solutions. The details of the decoupling limit can be found in \cite{Giusto:2012gt}. The charges measured at infinity and the local charges are related by
\begin{gather}
N_1 = \bar N_1  + n_2 n_3, \qquad N_2 = \bar N_2 + n_1 n_3, \nn
N_3 = \bar N_3 + n_1 n_2 - n_4^2, \qquad N_4 = \bar N_4 + n_3 n_4.
\end{gather}
Quantization of flux again allows us to take $N_i,\bar N_i,n_i \in \mathbb Z$ and the absence of CTCs constrains
\begin{gather}
n_i, \bar N_i \ge 0,  \qquad i=1,2,3  \label{NoCTCNewBR1} \\
n_1 n_2 > n_4^2.   \label{NoCTCNewBR2} 
\end{gather}
From the additional constraints given in \cite{Giusto:2012gt} to avoid CTCs the relevant ones in the decoupling limit are
\begin{gather}
n_1 \bar N_1 + n_2 \bar N_2 -2 n_4 \bar N_4 \ge 0,   \label{NoCTCNewBR3} \\
\bar N_1 \bar N_2 - \bar N_4^2 \ge 0.  \label{NoCTCNewBR4}
\end{gather}
The entropy of the black ring is given by
\be
S=2 \pi \sqrt{D} 
\ee
where
\bea
D &=&n_1 n_2( N_1 N_2 - n_3(|J_L| - n_3 N_3 - n_3n_4^2 + 2 n_4 N_4)) -\f{1}{4} (|J_L| -J_R -2 n_3  N_3 - 2n_3  n_4^2+ 2 n_4 N_4)^2 \nn
&& ~~~~~~~~~+ n_4 ( |J_L| -J_R -2 n_3  N_3 - 2n_3  n_4^2+ 2 n_4 N_4)  ( N_4 - n_3 n_4) - n_4^2 (N_1 - n_2 n_3) ( N_2 - n_1 n_3)  \nn
&& ~~~~~~~~~ - n_1 n_2 ( N_4 - n_3 n_4)^2 +  (n_1 +n_2 + n_3 - 2n_4) n_3 n_4^2. \label{NewBREntropy}
\eea
In addition we have the constraint
\be
|J_L| - J_R  = \bar N_1 n_1 + \bar N_2 n_2 + \bar N_3 n_3 + 2 n_1 n_2 n_3 - 2 n_4 N_4 \label{Eq:NewBR-contraint}
\ee
relating the two angular momenta of  the black ring.
We again consider $N_1 =N_2 = \sqrt{N}$ and $n_1 = n_2 = \sqrt{n}$ for simplicity. Since the net $NS1$ and $NS5$ charges are zero for the CFT phase we limit our considerations to $N_4=0$. This immediately takes care of the constraint \bref{NoCTCNewBR3}. We first analyze the existence of the black ring for $J_L>0$ ($k \ge 0$) and then use $J_L \to - J_L$ symmetry of \bref{NewBREntropy} to infer the existence for $J_L<0$.  If we take the black ring to be dual to the new phase in the $k^{th}$ wedge we  get the  additional constraint 
\be
J_R =J_L(2k+1)  - 2 N_p-2 k (k+1)N \label{JRKthWedge}
\ee
from spectral flowing the $0^{th}$ wedge CFT constraint \bref{Eqn:JRconstraint} to the $k^{th}$ wedge. 
Inserting \bref{JRKthWedge} in \bref{Eq:NewBR-contraint} and solving for $\sqrt{n}$ we get
\be
\sqrt{n} = \frac{\sqrt{N}-\sqrt{2 n_3 k  ( J_L - (k+1) N)
 +N+n_3^2 N_3-2
   n_3 N_3+  n_3^2 n_4^2}}{n_3}
\ee
where the other root is dropped to avoid CTCs as per the conditions in  \bref{NoCTCNewBR1}. Furthermore \bref{NoCTCNewBR2} requires 
\be
n > n_4^2
\ee
which amounts to
\bea
(2-n_3) N_3&>& 2 k (J_L - (k+1) N)   + 2 n_4 \sqrt{N} \label{NewBRNoCTC} 
\eea
 \begin{figure}[t]
 \begin{center}
\begin{tabular}{ccc}
 \epsfxsize=5cm \epsfbox{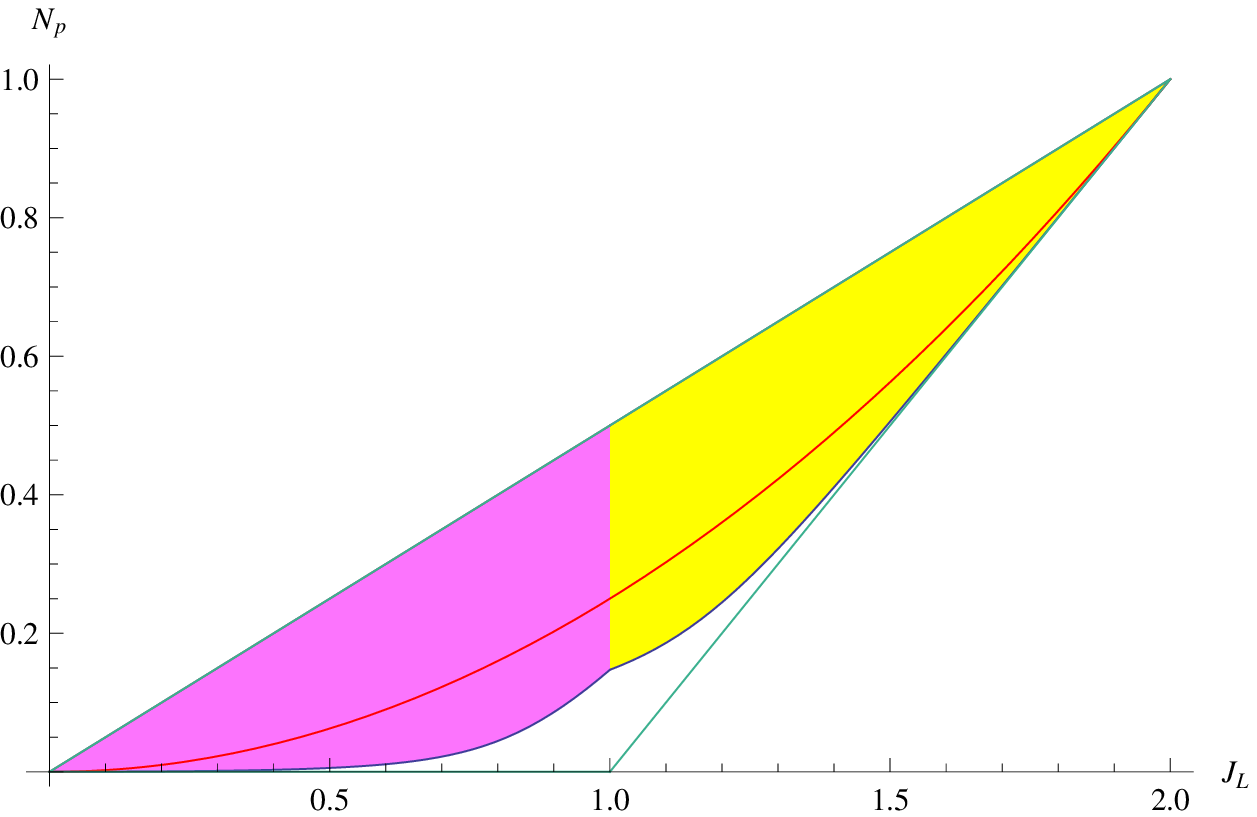} 
 &
 \epsfxsize=5cm \epsfbox{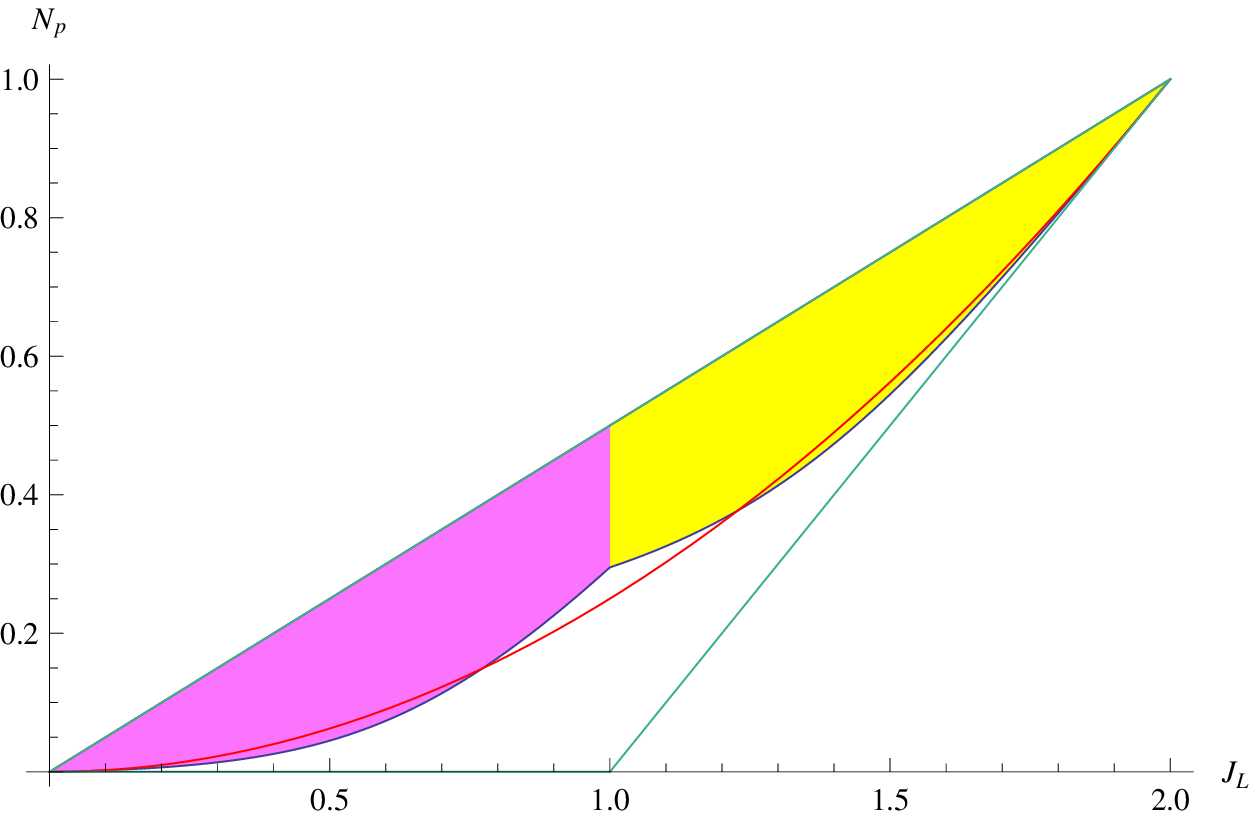} 
 &
 \epsfxsize=5cm \epsfbox{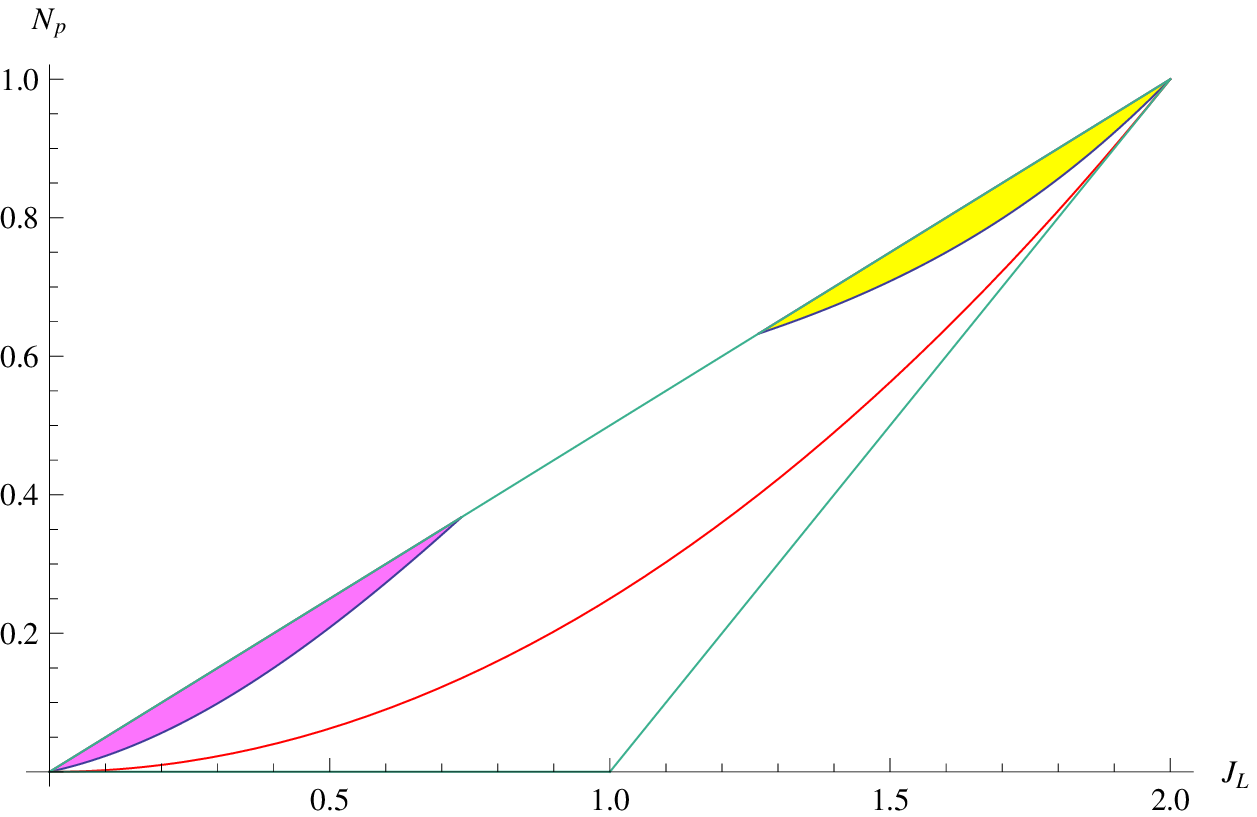} \\
(a) ~ $n_4 = \sqrt{.001 N}$ & (b) ~  $n_4 = \sqrt{.01 N}$ & (c) ~  $n_4 = \sqrt{.1 N}$ \\ 
\end{tabular}
\caption{\sl The new phase for various values of $n_4$. The green lines are the unitarity bound. The red line is the cosmic censorship bound for a single center BMPV black hole. The yellow region is the black ring and the pink region is spectral flow of the black ring after taking $J_L \to -J_L$. For $n_4=0$ this is a supertube and BMPV bound state. \label{fig:newPhaseWithNewDipole}}
\end{center}
\vspace*{-.5cm}
\end{figure}
Since this condition is stronger than \bref{STUNoCTC} we still get the condition that the new black ring can only exist in the wedges $0,1,-1,-2$. As in the previous section we can evaluate the entropies in the four wedges and spectral flow to the $0^{th}$ wedge. We get
\begin{gather}
D_0 =  (N_p - J_L + N)( \sqrt{N} - \sqrt{N- N_p+n_4^2})^2 - n_4^2(N-N_p) \nn
+ n_4^2 ( 2(  \sqrt{N} - \sqrt{N- N_p+n_4^2})+1-2n_4),\\
D_{-1 \to 0}  = N_p (\sqrt{N} - \sqrt{J_L - N_p+n_4^2})^2 -n_4^2 (J_L-N_p) \nn
+n_4^2(2(\sqrt{N} - \sqrt{J_L - N_p+n_4^2})+1-2n_4), \\
D_{1 \to 0} = D_{-1 \to 0}  - J_L^2,  \\
D_{-2 \to 0} =  D_0 - (J_L-2N)^2. 
\end{gather}
As in the previous section we have written $N_3$ as $N_p$ to emphasize that the third charge is momentum in the type IIB frame.

Thus clearly we have to only consider $D_0$ and $D_{-1 \to 0}$. In the $0^{th}$ wedge the condition \bref{NoCTCNewBR4} becomes $N_p \le N$ and is clearly satisfied. Since the $-1^{th}$wedge  is related to $0^{th}$ wedge by $J_L \to - J_L$ the no CTC condition is satisfied in that wedge as well.
We see that for $n_4 \sim O(1)$ the first term scales with $O(N^2)$ and the rest are subleading so can be dropped. It is clear that in this regime the entropy is maximized for $n_4=0$. For $n_4 \sim O(\sqrt{N})$ the first two terms are of the same order and the rest subleading. In this regime the entropy formulae become
\begin{gather}
D_0 =  (N_p - J_L + N)( \sqrt{N} - \sqrt{N- N_p+n_4^2})^2 - n_4^2(N-N_p), \\
D_{-1 \to 0}  = N_p (\sqrt{N} - \sqrt{J_L - N_p+n_4^2})^2 -n_4^2 (J_L-N_p). 
\end{gather}
and it is again clear that the entropy is maximized for $n_4=0$. From the no CTC condition \bref{NewBRNoCTC} we get $N_p > 2 n_4 \sqrt{N}$ for the black ring in the $0^{th}$ wedge and $N_p - J_L + N > 2 n_4 \sqrt{N}$ for the spectral flowed black ring from $-1^{th}$ wedge. Along with $N_p, ~J_L \sim O(N)$ this ensures that $O(\sqrt{N})$ is the maximum amount for the fourth dipole charge.
 
We will now plot the gravity phases with non-zero $n_4$. The phase only exists for $D_0$ or $D_{-1 \to 0}$ greater than zero. It turns out that for $N_p = 2 n_4 \sqrt{N}$ we have $D_0<0$ so we do not have to worry about this bound when plotting the new phase. Likewise, for $N_p - J_L + N = 2 n_4 \sqrt{N}$  we have $D_{-1 \to 0}<0$ so this bound does not matter. In figure \ref{fig:newPhaseWithNewDipole} we plot the bulk solution for $n_4= \sqrt{.001 N},\sqrt{.01 N },\sqrt{.1 N}$. The yellow region is the new black ring and the pink is spectral flow away from a black ring from the $-1^{th}$ wedge.  For $n_4=0$ this was a bound state of BMPV and supertube. It should be possible to generalize the spectral flow from \cite{Bena:2008wt} to include the fourth charge and dipole charge to see what this phase is explicitly. However, since it is of no immediate relevance to us, we do not do so here. It is interesting to note that the configurations with $n_4 \ne 0$ continue to exist outside the cosmic censorship bound for small values of $n_4$. Additionally, it is easy to see that these configurations are entropically dominant over the BMPV in a some region for small $n_4$ so they are also examples of entropy enigma.

\section{Discussion} \label{Conclusion}

In this paper we first reviewed  \cite{Bena:2011zw} where supersymmetric phases of the D1-D5 system were studied on both sides of the AdS/CFT correspondence. Outside the Cardy regime the orbifold point had a new entropically dominant phase which comes into existence for  $J_L>2 N_p$ and existed all the way to the unitarity bound. This phase had $J_R=J_L-2N_p$. The bulk configurations considered were within the STU truncation and were taken to have $J_R=J_L-2N_p$ also. These configurations had less entropy than the CFT phase but in a smaller region were still entropically dominant over the single center BMPV black hole. These can be thought of as the BMPV shedding off excess charges as hair, or moulting. For $J_L<N$ the BMPV sheds a supertube and for $J_L>N$ it sheds hair of Taub-NUT charge and becomes a black ring. These configurations are related to each other by spectral flow. These phase diagrams are plotted in figure \ref{fig:new_phase_diag}. 

In the second part of the paper we discuss whether the new black rings found in \cite{Giusto:2012gt}, which have an additional dipole charge compared to the STU black rings studied in  \cite{Bena:2011zw}, could account for the missing entropy. As before we focussed on black rings, and configurations related by spectral flow, with the constraint $J_R=J_L-2 N_p$ and found that the extra dipole charge always lowers the entropy. Since the entropy of the STU black ring went to zero at the unitarity bound, this means that extra dipole charge causes the region of existence to shrink away from the unitarity bound. As a by product this shows us that it is not possible to have any of the duals of the RR ground states carry any fourth dipole charge. This is consistent with \cite{Rychkov:2005ji} where the bulk configurations of \cite{Lunin:2001fv}, which did not have any fourth dipole, were geometrically quantized to reproduce the CFT entropy of Ramond-Ramond ground states.

The resulting phase digram is shown for various values of the dipole charge in figure \ref{fig:newPhaseWithNewDipole}. The yellow region is the black ring and the pink is a spectral flow of it after taking $J_L \to - J_L$. In the absence of the fourth dipole charge the pink configuration would be a bound state of a supertube and BMPV. 

It would be interesting to perform the explicit spectral flow by generalizing the rules in \cite{Bena:2008wt} to see what this generalized bound state of a supertube and BMPV is. However, for now we just discuss this qualitatively. Since solutions with different values of the fourth dipole charge are continuously connected for large $N$, we expect the configuration to be a bound state of a black hole and a supertube {\em like}  smooth center. Since the supertube is one of the RR ground states and we have seen that it is not possible to turn on the fourth dipole charge alone for this state it is likely that the smooth center will have some fourth charge and may also have some dipole charge\footnote{Such supertubes with the fourth charge but without the fourth dipole charge have recently been constructed in \cite{Vasilakis:2012zg}.}. The black hole in the center cannot have a dipole charge but it might have the fourth charge in such a way that the net fourth charge of the bound state is zero. Note that since at least one, possibly two, of the centers has the fourth charge,  they cannot fit in the D1-D5 CFT independently but only in the modified CFT of \cite{Giusto:2012gt} with central charge
 \be
 c=6 (N_1 N_5 - N_4^2). \label{ModifiedCentralCharge}
 \ee

Although some work has been done on the CFT understanding of black rings and their dipoles \cite{Bena:2004tk,Iizuka:2005uv,Dabholkar:2005qs,Alday:2005xj,Dabholkar:2006za} our current understanding is far from satisfactory. It would be interesting to see what constraints may be put on the CFT phase of section \ref{ReviewCFT} to account for the fourth dipole. It would also be interesting to see if there is an IR CFT which explains the entropy of these black rings along the lines of \cite{Cyrier:2004hj} although this approach does not shed light on interesting features of the UV CFT like dipoles and multi-center configurations.

Finally, it would be interesting to see if the black holes in the CFT with central charge \bref{ModifiedCentralCharge} also exhibits molting. While it seems reasonable that this would happen, it would be interesting to see if the resulting centers sit inside the same CFT (i.e.  have the same ratio of $N_4^2$ to $N_1 N_5$) as the parent black hole. This analysis would also shed light on the structure of the new CFT in \cite{Giusto:2012gt}  outside the Cardy regime.

\section*{Acknowledgments}

I thank Iosif Bena, Jan de Boer, Sheer El-Showk, Stefano Giusto, Rodolfo Russo, Masaki Shigemori and Orestis Vasilakis for
helpful discussions. My work is supported by the ERC
Advanced Grant 268088-EMERGRAV. 

\bibliographystyle{toine}
\bibliography{Papers.bib}

\providecommand{\href}[2]{#2}\begingroup\raggedright\begin{thebibliography}{10}

\bibitem{Bena:2011zw}
I.~Bena, B.~D. Chowdhury, J.~de~Boer, S.~El-Showk  and M.~Shigemori,
  \emph{{Moulting Black Holes}}, JHEP {\bf 1203} (2012) 094,
\href{http://www.arXiv.org/abs/1108.0411}{{\tt 1108.0411}}

\bibitem{Giusto:2012gt}
S.~Giusto and R.~Russo, \emph{{Adding new hair to the 3-charge black ring}},
  Class.Quant.Grav. {\bf 29} (2012) 085006,
  \href{http://www.arXiv.org/abs/1201.2585}{{\tt 1201.2585}},
18 pages

\bibitem{Gubser:2008px}
S.~S. Gubser, \emph{{Breaking an Abelian gauge symmetry near a black hole
  horizon}}, Phys.Rev. {\bf D78} (2008) 065034,
\href{http://www.arXiv.org/abs/0801.2977}{{\tt 0801.2977}}

\bibitem{Hartnoll:2008vx}
S.~A. Hartnoll, C.~P. Herzog  and G.~T. Horowitz, \emph{{Building a Holographic
  Superconductor}}, Phys.Rev.Lett. {\bf 101} (2008) 031601,
\href{http://www.arXiv.org/abs/0803.3295}{{\tt 0803.3295}}

\bibitem{Hartnoll:2008kx}
S.~A. Hartnoll, C.~P. Herzog  and G.~T. Horowitz, \emph{{Holographic
  Superconductors}}, JHEP {\bf 0812} (2008) 015,
\href{http://www.arXiv.org/abs/0810.1563}{{\tt 0810.1563}}

\bibitem{Denef:2009tp}
F.~Denef and S.~A. Hartnoll, \emph{{Landscape of superconducting membranes}},
  Phys.Rev. {\bf D79} (2009) 126008,
\href{http://www.arXiv.org/abs/0901.1160}{{\tt 0901.1160}}

\bibitem{Gubser:2009qm}
S.~S. Gubser, C.~P. Herzog, S.~S. Pufu  and T.~Tesileanu,
  \emph{{Superconductors from Superstrings}}, Phys.Rev.Lett. {\bf 103} (2009)
  141601,
\href{http://www.arXiv.org/abs/0907.3510}{{\tt 0907.3510}}

\bibitem{Gauntlett:2009dn}
J.~P. Gauntlett, J.~Sonner  and T.~Wiseman, \emph{{Holographic
  superconductivity in M-Theory}}, Phys.Rev.Lett. {\bf 103} (2009) 151601,
\href{http://www.arXiv.org/abs/0907.3796}{{\tt 0907.3796}}

\bibitem{Gauntlett:2009bh}
J.~P. Gauntlett, J.~Sonner  and T.~Wiseman, \emph{{Quantum Criticality and
  Holographic Superconductors in M-theory}}, JHEP {\bf 1002} (2010) 060,
\href{http://www.arXiv.org/abs/0912.0512}{{\tt 0912.0512}}

\bibitem{Correa:2010hf}
F.~Correa, C.~Martinez  and R.~Troncoso, \emph{{Scalar solitons and the
  microscopic entropy of hairy black holes in three dimensions}}, JHEP {\bf
  1101} (2011) 034,
\href{http://www.arXiv.org/abs/1010.1259}{{\tt 1010.1259}}

\bibitem{Basu:2010uz}
P.~Basu, J.~Bhattacharya, S.~Bhattacharyya, R.~Loganayagam, S.~Minwalla  {\em
  et al.}, \emph{{Small Hairy Black Holes in Global AdS Spacetime}}, JHEP {\bf
  1010} (2010) 045, \href{http://www.arXiv.org/abs/1003.3232}{{\tt 1003.3232}},
68+1 pages, 18 figures, JHEP format. v2 : small typos corrected and a reference
  added

\bibitem{Bhattacharyya:2010yg}
S.~Bhattacharyya, S.~Minwalla  and K.~Papadodimas, \emph{{Small Hairy Black
  Holes in $AdS_5 x S^5$}}, JHEP {\bf 1111} (2011) 035,
\href{http://www.arXiv.org/abs/1005.1287}{{\tt 1005.1287}}

\bibitem{Dias:2011tj}
O.~J. Dias, P.~Figueras, S.~Minwalla, P.~Mitra, R.~Monteiro  {\em et al.},
  \emph{{Hairy black holes and solitons in global AdS 5}},
\href{http://www.arXiv.org/abs/1112.4447}{{\tt 1112.4447}}

\bibitem{Alsup:2011qn}
J.~Alsup, G.~Siopsis  and J.~Therrien, \emph{{Hair on near-extremal
  Reissner-Nordstrom AdS black holes}},
\href{http://www.arXiv.org/abs/1110.3342}{{\tt 1110.3342}}

\bibitem{Anabalon:2012sn}
A.~Anabalon, F.~Canfora, A.~Giacomini  and J.~Oliva, \emph{{Black Holes with
  Primary Hair in gauged N=8 Supergravity}},
  \href{http://www.arXiv.org/abs/1203.6627}{{\tt 1203.6627}},
9 pages, no figures

\bibitem{Brihaye:2012cb}
Y.~Brihaye and B.~Hartmann, \emph{{Hairy charged Gauss-Bonnet solitons and
  black holes}}, \href{http://www.arXiv.org/abs/1203.3109}{{\tt 1203.3109}},
20 pages inlcuding 17 figures

\bibitem{Capela:2012uk}
F.~Capela and G.~Nardini, \emph{{Hairy Black Holes in Massive Gravity:
  Thermodynamics and Phase Structure}},
\href{http://www.arXiv.org/abs/1203.4222}{{\tt 1203.4222}}

\bibitem{Chowdhury:2010ct}
B.~D. Chowdhury and A.~Virmani, \emph{{Modave Lectures on Fuzzballs and
  Emission from the D1-D5 System}},
\href{http://www.arXiv.org/abs/1001.1444}{{\tt 1001.1444}}

\bibitem{Gauntlett:2004wh}
J.~P. Gauntlett and J.~B. Gutowski, \emph{{Concentric black rings}}, Phys.Rev.
  {\bf D71} (2005) 025013,
\href{http://www.arXiv.org/abs/hep-th/0408010}{{\tt hep-th/0408010}}

\bibitem{Denef:2007vg}
F.~Denef and G.~W. Moore, \emph{{Split states, entropy enigmas, holes and
  halos}}, JHEP {\bf 1111} (2011) 129,
  \href{http://www.arXiv.org/abs/hep-th/0702146}{{\tt hep-th/0702146}},
149 pages, 21 figures

\bibitem{deBoer:2008fk}
J.~de~Boer, F.~Denef, S.~El-Showk, I.~Messamah  and D.~Van~den Bleeken,
  \emph{{Black hole bound states in AdS(3) x S**2}}, JHEP {\bf 0811} (2008)
  050,
\href{http://www.arXiv.org/abs/0802.2257}{{\tt 0802.2257}}

\bibitem{Anninos:2011vn}
D.~Anninos, T.~Anous, J.~Barandes, F.~Denef  and B.~Gaasbeek, \emph{{Hot Halos
  and Galactic Glasses}}, JHEP {\bf 1201} (2012) 003,
\href{http://www.arXiv.org/abs/1108.5821}{{\tt 1108.5821}}

\bibitem{Chowdhury:2011qu}
B.~D. Chowdhury and B.~Vercnocke, \emph{{New instability of non-extremal black
  holes: spitting out supertubes}}, JHEP {\bf 1202} (2012) 116,
\href{http://www.arXiv.org/abs/1110.5641}{{\tt 1110.5641}}

\bibitem{Breckenridge:1996is}
J.~Breckenridge, R.~C. Myers, A.~Peet  and C.~Vafa, \emph{{D-branes and
  spinning black holes}}, Phys.Lett. {\bf B391} (1997) 93--98,
\href{http://www.arXiv.org/abs/hep-th/9602065}{{\tt hep-th/9602065}}

\bibitem{Seiberg:1999xz}
N.~Seiberg and E.~Witten, \emph{{The D1 / D5 system and singular CFT}}, JHEP
  {\bf 9904} (1999) 017,
\href{http://www.arXiv.org/abs/hep-th/9903224}{{\tt hep-th/9903224}}

\bibitem{deBoer:1998us}
J.~de~Boer, \emph{{Large N elliptic genus and AdS / CFT correspondence}}, JHEP
  {\bf 9905} (1999) 017,
\href{http://www.arXiv.org/abs/hep-th/9812240}{{\tt hep-th/9812240}}

\bibitem{Maldacena:1999bp}
J.~M. Maldacena, G.~W. Moore  and A.~Strominger, \emph{{Counting BPS black
  holes in toroidal Type II string theory}},
\href{http://www.arXiv.org/abs/hep-th/9903163}{{\tt hep-th/9903163}}

\bibitem{Bena:2008wt}
I.~Bena, N.~Bobev  and N.~P. Warner, \emph{{Spectral Flow, and the Spectrum of
  Multi-Center Solutions}}, Phys.Rev. {\bf D77} (2008) 125025,
\href{http://www.arXiv.org/abs/0803.1203}{{\tt 0803.1203}}

\bibitem{Dabholkar:2009dq}
A.~Dabholkar, M.~Guica, S.~Murthy  and S.~Nampuri, \emph{{No entropy enigmas
  for N=4 dyons}}, JHEP {\bf 1006} (2010) 007,
\href{http://www.arXiv.org/abs/0903.2481}{{\tt 0903.2481}}

\bibitem{Dhar:1999ax}
A.~Dhar, G.~Mandal, S.~R. Wadia  and K.~Yogendran, \emph{{D-1 / D-5 system with
  B field, noncommutative geometry and the CFT of the Higgs branch}},
  Nucl.Phys. {\bf B575} (2000) 177--194,
\href{http://www.arXiv.org/abs/hep-th/9910194}{{\tt hep-th/9910194}}

\bibitem{Giusto:2011fy}
S.~Giusto, R.~Russo  and D.~Turton, \emph{{New D1-D5-P geometries from string
  amplitudes}}, JHEP {\bf 1111} (2011) 062,
\href{http://www.arXiv.org/abs/1108.6331}{{\tt 1108.6331}}

\bibitem{Elvang:2004rt}
H.~Elvang, R.~Emparan, D.~Mateos  and H.~S. Reall, \emph{{A Supersymmetric
  black ring}}, Phys.Rev.Lett. {\bf 93} (2004) 211302,
\href{http://www.arXiv.org/abs/hep-th/0407065}{{\tt hep-th/0407065}}

\bibitem{Avery:2009tu}
S.~G. Avery, B.~D. Chowdhury  and S.~D. Mathur, \emph{{Emission from the D1D5
  CFT}}, JHEP {\bf 10} (2009) 065,
\href{http://www.arXiv.org/abs/0906.2015}{{\tt 0906.2015}}

\bibitem{David:2002wn}
J.~R. David, G.~Mandal  and S.~R. Wadia, \emph{{Microscopic formulation of
  black holes in string theory}}, Phys.Rept. {\bf 369} (2002) 549--686,
\href{http://www.arXiv.org/abs/hep-th/0203048}{{\tt hep-th/0203048}}

\bibitem{Bena:2007kg}
I.~Bena and N.~P. Warner, \emph{{Black holes, black rings and their
  microstates}}, Lect. Notes Phys. {\bf 755} (2008) 1--92,
\href{http://www.arXiv.org/abs/hep-th/0701216}{{\tt hep-th/0701216}}

\bibitem{Rychkov:2005ji}
V.~S. Rychkov, \emph{{D1-D5 black hole microstate counting from supergravity}},
  JHEP {\bf 0601} (2006) 063,
\href{http://www.arXiv.org/abs/hep-th/0512053}{{\tt hep-th/0512053}}

\bibitem{Lunin:2001fv}
O.~Lunin and S.~D. Mathur, \emph{{Metric of the multiply wound rotating
  string}}, Nucl.Phys. {\bf B610} (2001) 49--76,
\href{http://www.arXiv.org/abs/hep-th/0105136}{{\tt hep-th/0105136}}

\bibitem{Vasilakis:2012zg}
O.~Vasilakis, \emph{{Bubbling the Newly Grown Black Ring Hair}},
  \href{http://www.arXiv.org/abs/1202.1819}{{\tt 1202.1819}},
13 pages

\bibitem{Bena:2004tk}
I.~Bena and P.~Kraus, \emph{{Microscopic description of black rings in AdS /
  CFT}}, JHEP {\bf 0412} (2004) 070,
\href{http://www.arXiv.org/abs/hep-th/0408186}{{\tt hep-th/0408186}}

\bibitem{Iizuka:2005uv}
N.~Iizuka and M.~Shigemori, \emph{{A Note on D1-D5-J system and 5-D small black
  ring}}, JHEP {\bf 0508} (2005) 100,
\href{http://www.arXiv.org/abs/hep-th/0506215}{{\tt hep-th/0506215}}

\bibitem{Dabholkar:2005qs}
A.~Dabholkar, N.~Iizuka, A.~Iqubal  and M.~Shigemori, \emph{{Precision
  microstate counting of small black rings}}, Phys.Rev.Lett. {\bf 96} (2006)
  071601,
\href{http://www.arXiv.org/abs/hep-th/0511120}{{\tt hep-th/0511120}}

\bibitem{Alday:2005xj}
L.~F. Alday, J.~de~Boer  and I.~Messamah, \emph{{What is the dual of a
  dipole?}}, Nucl.Phys. {\bf B746} (2006) 29--57,
\href{http://www.arXiv.org/abs/hep-th/0511246}{{\tt hep-th/0511246}}

\bibitem{Dabholkar:2006za}
A.~Dabholkar, N.~Iizuka, A.~Iqubal, A.~Sen  and M.~Shigemori, \emph{{Spinning
  strings as small black rings}}, JHEP {\bf 0704} (2007) 017,
\href{http://www.arXiv.org/abs/hep-th/0611166}{{\tt hep-th/0611166}}

\bibitem{Cyrier:2004hj}
M.~Cyrier, M.~Guica, D.~Mateos  and A.~Strominger, \emph{{Microscopic entropy
  of the black ring}}, Phys.Rev.Lett. {\bf 94} (2005) 191601,
\href{http://www.arXiv.org/abs/hep-th/0411187}{{\tt hep-th/0411187}}

\end{thebibliography}\endgroup

\end{document}